\documentclass[a4paper]{article}

\usepackage{INTERSPEECH2022}
\usepackage{amsfonts}
\usepackage{multirow}
\usepackage{bbding}
\usepackage{cite}
\usepackage[urlcolor=blue]{hyperref}
\usepackage{subfigure}
\usepackage{mathrsfs}

\def\model{FS-CANet}
\def\camodule{FSCA}

\title{Speech Enhancement with Fullband-Subband Cross-Attention Network}
\name{Author Name$^1$, Co-author Name$^2$}
\name{Jun Chen$^{1,\dagger}$\thanks{$^{\dagger}$ Work conducted when the first author was intern at Tencent.}, Wei Rao$^{2,*}$, Zilin Wang$^1$, Zhiyong Wu$^{1,3,*}$\thanks{$^{*}$ Corresponding author.}, Yannan Wang$^2$, \\ Tao Yu$^2$, Shidong Shang$^2$, Helen Meng$^{1,3}$}
\address{
  $^1$Shenzhen International Graduate School, Tsinghua University, Shenzhen, China\\
  $^2$Tencent Ethereal Audio Lab, Shenzhen, China \\
  $^3$The Chinese University of Hong Kong, Hong Kong SAR, China}
\email{y-chen21@mails.tsinghua.edu.cn, ellenwrao@tencent.com, zywu@sz.tsinghua.edu.cn}

\begin{document}
\maketitle 
\begin{abstract}
FullSubNet has shown its promising performance on speech enhancement by utilizing both fullband and subband information. 
However, the relationship between fullband and subband in FullSubNet is achieved by simply concatenating the output of fullband model and subband units. 
It only supplements the subband units with a small quantity of global information and has not considered the interaction between fullband and subband. 
This paper proposes a fullband-subband cross-attention (\camodule{}) module to interactively fuse the global and local information and applies it to FullSubNet. 
This new framework is called as \model{}. Moreover, different from FullSubNet, the proposed \model{} optimize the fullband extractor by temporal convolutional network (TCN) blocks to further reduce the model size. 
Experimental results on DNS Challenge - Interspeech 2021 dataset show that the proposed \model{} outperforms other state-of-the-art speech enhancement approaches, and demonstrate the effectiveness of fullband-subband cross-attention.


\end{abstract}
\noindent\textbf{Index Terms}: interactive fusion, fullband-subband cross-attention, speech enhancement
\section{Introduction}

The interference of environmental noise is one of the main factors that hinder the speech communication.
Single-channel speech enhancement methods remove background noise from single-channel noisy audio signals, aiming to improve the quality and intelligibility of the speech, and have significant applications in hearing aids, audio communication and automatic speech recognition.
Traditional speech enhancement methods use statistical signal theory to effectively suppress stationary noise, but they do not perform well under conditions of low signal-to-noise ratio (SNR).
In the past few years, deep learning-based methods have achieved promising results, especially in dealing with non-stationary noise in complex acoustic environments.
The neural networks can enhance the noisy speech either in frequency-domain or directly in time-domain.
The frequency-domain approaches \cite{2014regression, wang2018supervised, chen2017long, erdogan2015phase} generally take the noisy spectral feature as the input,
and their learning target is the clean spectral feature or the mask such as Ideal Ratio Mask\cite{irm} and complex Ideal Ratio Mask (cIRM)\cite{cirm}.
On the other hand, The time-domain approaches \cite{2018wavenet, 2018waveunet, 2019tcnn} predict a clean speech waveform from the corresponding noisy speech waveform.
Overall, the frequency domain approaches is more preferable considering the robustness of system and computational complexity \cite{yin2020phasen}.

The subband model \cite{2020subband} is a previously proposed work for frequency-domain speech enhancement. 
The work is performed in a subband style: the input of the model consists of one frequency together with several contextual frequencies. 
Thus the subband model learns the frequency-wise signal stationarity to discriminate between speech and stationary noise.
However, since it can not model the global spectral pattern and exploit the long distance cross-band dependencies, it is difficult to recover clean speech on subband with low SNR.
To solve this problem, the FullSubNet \cite{hao2021fullsubnet} introduces a fullband model extracting the global spectral information on the basis of the subband model, and performs joint optimization after connecting the two in series. 
By means of this way, the FullSubNet can capture the global spectral context \cite{2014regression, wang2018supervised} while retaining the ability to model signal stationarity and attend the local spectral patterns.
Consequently, FullSubNet achieves excellent results on DNS Challenge dataset \cite{dnschallenge}.

The success of FullSubNet effectively demonstrates the importance of global spectral information for the subband model.
However, in FullSubNet, the relationship between fullband model and subband model is achieved by simply concatenating the output of fullband model and subband units. This concatenation method only supplements a small amount of global information for the subband units. It lacks the interaction between fullband information and subband information, which limits the potential of the FullSubNet.

To address the above issue of FullSubNet, this paper proposes the fullband-subband cross-attention (\camodule{}) module to interactively fuse the fullband and subband information and applies it to FullSubNet. Furthermore, a fullband extractor \cite{fullsubnetplus} composed of TCN blocks \cite{bai2018empirical} is used to instead the fullband model in FullSubNet for reducing model size. This new network is called as \model{}.
Experimental results on the DNS Challenge dataset show that our \model{} outperforms FullSubNet+ \cite{fullsubnetplus} and FullSubNet in terms of both number of parameters and performance.
Furthermore, the \model{} also exceeds other state-of-the-art speech enhancement methods on the DNS Challenge - interspeech 2021 dataset. These experimental results also demonstrate that the fullband-subband cross-attention is an effective interactive fusion method.

\begin{figure*}[!htbp]
	\centering
    \includegraphics[width=1.01\linewidth]{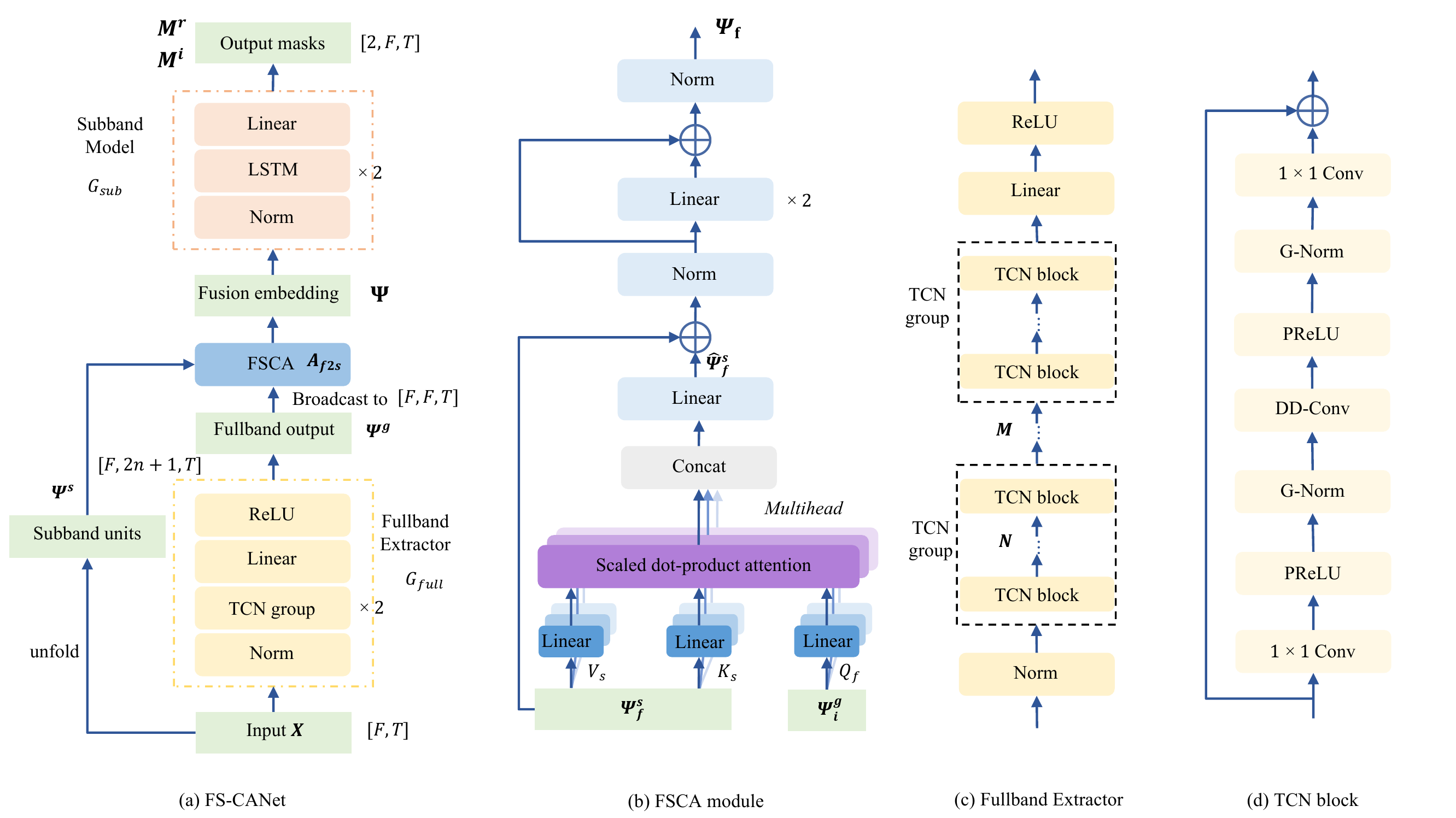}
    \vspace{-0.5cm}
	\caption{(a) The overall diagram of the proposed \model{}. The model mainly contains a Fullband Extractor, a \camodule{} module and a Subband Model.
	(b) The details of the \camodule{} module, where ``Multihead" refers to the multi-head attention.
	(c) The details of the Fullband Extractor.
	(d) The details of TCN block. The ``DD-Conv" indicates a dilated depth-wise separable convolution. The ``G-norm" is a global layer normalization \cite{luo2019conv}.}
	\label{fig:totalarch}
	 \vspace{-0.3cm}
\end{figure*}

\section{\model{}}


This paper only focuses on the denoising task in the STFT domain, and the target is to suppress noise and recover the reverberant speech signal (the reverberant signal received at the microphone). \model{} is proposed to interactively fuse the fullband and subband information by fullband-subband cross-attention module.

The architecture of \model{} is shown in Fig.\ref{fig:totalarch}(a), which mainly consists of three main parts: a fullband extractor $G_{full}$, a \camodule{} module $A_{f2s}$, and a subband model $G_{sub}$.
To begin with, the input $\mathbf{X}  \in \mathbb{R}^{F \times T}$ is fed to the fullband extractor $G_{full}$, and additionally it is also the subband units $\mathbf{\Psi}^{s}$ after unfolding.
Then the fullband extractor $G_{full}$ extracts the global spectral information from the input spectrogram and outputs the fullband embedding $\mathbf{\Psi}^{g}$. 
After a python broadcast, $\mathbf{\Psi}^{g}$ is treated as the input of \camodule{} module $A_{f2s}$ together with the subband units $\mathbf{\Psi}^{s}$.
Next, $A_{f2s}$ obtains the fusion embeddings $\mathbf{\Psi}$ after interactively fusing $\mathbf{\Psi}^{s}$ and $\mathbf{\Psi}^{g}$.
Finally, taking the $\mathbf{\Psi}$ as input, the subband model $G_{sub}$ predicts the learning target cIRM $\mathbf{M}^{r}$ and $\mathbf{M}^{i}$.
In the following these modules are described in detail.



\subsection{Fullband Extractor}
\label{sec:2.1}
The fullband extractor is an efficient fullband processing model, which is illustrated in \cite{fullsubnetplus}.
Compared with the fullband model in FullSubNet, it has a better performance with a smaller number of parameters \cite{fullsubnetplus}.
Therefore, the fullband extractor is used to replace the original fullband model in FullSubNet.

As a model with powerful temporal sequence modeling ability, TCN blocks have been widely used in speech separation \cite{luo2019conv, fan2020end, tzinis2020two}
and target speaker extraction tasks \cite{xu2020spex} recently.
Fig.\ref{fig:totalarch}(d) shows the structure of the TCN block, which consists of three main components, namely input $1 \times 1 \ Conv$, depthwise dilated convolution ($DD-Conv$) and output $1 \times 1 \  Conv$.
Parametric ReLU (PReLU) activation function and normalization layers are inserted between adjacent convolutions.
Residual connection is applied to alleviate gradient vanishing problem.
In fullband extractor, similar to Conv-TasNet \cite{luo2019conv}, the TCN blocks are stacked by exponentially increasing the dilation factor in each group to capture the features of speech signals with long-range dependence over the fullband.
As shown in Fig.\ref{fig:totalarch}(c), for the fullband extractor, there are $M$ groups of $N$ TCN blocks, where $M$ and $N$ are hyper-parameters. 
A fully connected layer and a ReLU activation function are deployed after these stacked TCN blocks.

The fullband extractor extracts fullband information and outputs the fullband embedding $\mathbf{\Psi}^{g}$ with the same size as its input $\mathbf{X}$, which is expected to provide complementary information for the subband units $\mathbf{\Psi}^{s}$.
The fullband embedding and subband units together serve as the input of $A_{f2s}$.

\subsection{Fullband-Subband Cross-Attention}
\label{sec:2.2}
Recently, the multi-head cross attention mechanism has shown a surprising potential in speech-related interactive fusion tasks like multi-modal active speaker detection \cite{tao2021someone} and speaker extraction \cite{wang2021neural}.
Inspired by this, \camodule{} module is proposed for interactively fusing the fullband and subband information to replace the simple concatenation \cite{hao2021fullsubnet} in original FullSubNet.
\subsubsection{The concatenation in FullSubNet}
\label{sec:2.2.1}
For each frequency $f$, we take a frequency bin vector $\mathbf{X}_{f} \in \mathbb{R}^{T}$ and the $2 \times n$ frequency bin vectors adjacent to it in the frequency domain from the weighted magnitude spectrogram $\mathbf{X}$ as a subband unit $\mathbf{\Psi}^{s}_{f}$:
\begin{equation}
\begin{split}
\mathbf{\Psi}^{s}_{f} =& [\mathbf{X}_{f-n}, \cdots ,\mathbf{X}_{f}, \cdots , \mathbf{X}_{f+n}] \in \mathbb{R}^{(2n+1) \times T}.
\end{split}
\end{equation}
In addition, circular Fourier frequencies are used for boundary frequencies.

According to the concatenation method of FullSubNet, each subband unit is concatenated with a frequency bin vector of the output of fullband extractor, denoted as $\mathbf{\Psi}^{g}_{f} \in \mathbb{R}^{T}$, to serve as the input $\widetilde{\mathbf{\Psi}}_{f}$ to the $G_{sub}$:
\begin{equation}
\begin{split}
\widetilde{\mathbf{\Psi}}_{f} = [\mathbf{\Psi}^{s}_{f}, \mathbf{\Psi}^{g}_{f}] \in \mathbb{R}^{(2n+2) \times T}.
\end{split}
\end{equation}
In this way, each subband unit is actually supplemented with the output of fullband extractor in only one frequency domain dimension, which does not make an adequate use of the global information.
Furthermore, it lacks the interaction of global and subband information.
To cope with the above issues, we propose the \camodule{} module.

\subsubsection{The details of Fullband-Subband Cross-Attention}
\label{sec:2.2.2}
Different from the existing work on frequency-domain self-attention \cite{dang2022dpt}, our \camodule{} module focuses on the interfusion of the output of the full-band model and the subband units to convey more fullband temporal and spectral information to the subband units.
The structure of the \camodule{} module $A_{f2s}$ is shown in Fig.\ref{fig:totalarch}(b), which mainly consists of the multi-headed attention layer for interactive fusion and linear layers for further fusion.
Besides, residual connections are applied in the \camodule{} module to alleviate the gradient vanishing problem.

The \camodule{} module $A_{f2s}$ takes the fullband embeddings and subband units together as the input.
Following section \ref{sec:2.2.1}, we get a total of $F$ subband units.
For the purpose of making each subband unit interact with a fullband embedding, we obtain $F$ fullband embeddings $\mathbf{\Psi}^{g}$ by python broadcast.
These subband units and fullband embeddings serve as the input to the $A_{f2s}$.
The \camodule{} module $A_{f2s}$ processes these $F$ pairs of subband units and fullband embeddings in parallel.
For each subband unit $\mathbf{\Psi}^{s}_{f}$ and fullband embedding $\mathbf{\Psi}^{g}$, the $\mathbf{\Psi}^{s}_{f}$ is linearly converted to $\mathbf{V}_s$, $\mathbf{K}_s \in \mathbb{R}^{F \times T}$ while the $\mathbf{\Psi}^{g}$ is linearly converted to $\mathbf{Q}_g  \in \mathbb{R}^{F \times T}$.
A multi-head attention mechanism is carried out to achieve the interaction between fullband and subband information:
\begin{equation}
\widehat{\mathbf{\Psi}}^{s}_{f} = Multihead(\mathbf{Q}_g, \mathbf{K}_s, \mathbf{V}_s)
\end{equation}
where $Multihead(\cdot)$ denotes the multi-head attention and the $\widehat{\mathbf{\Psi}}^{s}_{f}$ represents the output with the same size as $\mathbf{\Psi}^{s}_{f}$.
Through this interactive fusion, $\widehat{\mathbf{\Psi}}^{s}_{f}$ contains global temporal and spectral information at multiple dimensions that $\mathbf{\Psi}^{s}_{f}$ focus on.
Then, the $\widehat{\mathbf{\Psi}}^{s}_{f}$ is added to $\mathbf{\Psi}^{s}_{f}$.
Eventually, we stack another two linear layers with residual connection to further fuse them into the fusion embedding $\mathbf{\Psi}_{f} \in \mathbb{R}^{(2n+1) \times T}$.  
The total of $F$ fusion embeddings output by the \camodule{} module are then fed to the subband model $G_{sub}$.


\subsection{Subband Model}
\label{sec:2.3}
For the benefit of learning the frequency-wise signal stationarity while maintaining stability in model training, as shown in Fig.\ref{fig:totalarch}(a), the subband model $G_{sub}$ applies a structure composed of two stacked unidirectional LSTM layers and one fully connected layer instead of stacked TCN blocks.

We obtain a total of $F$ fusion embeddings after the interactive fusion of the \camodule{} module. 
Each fusion embedding contains the local spectral patterns as well as the global spectral information, both of which complement each other.
All of the fusion embeddings are fed into the subband model with shared parameters in parallel.
In subband model, the stacked LSTM layers learn the global and local frequency-wise signal stationarity based on these fusion embeddings.
Finally, the fully-connected layer outputs the cIRM as our learning target.

\begin{table*}[!htbp]
    \begin{center}
    \caption{The performance of WB-PESQ [MOS], NB-PESQ [MOS], STOI [\%], and SI-SDR [dB] on the DNS Challenge test dataset.} 
    \vspace{-0.3cm}
    \label{table:large total compare}
    \scalebox{1.0}{
    \begin{tabular}{cccccccccccc}
    \toprule
        \multirow{2}{*}{Model} & \multirow{2}{*}{Year}  & \multirow{2}{*}{\shortstack{\# Para\\(M)}}&  \multicolumn{4}{c}{With Reverb} & \multicolumn{4}{c}{Without Reverb} \\
    \cmidrule(lr){4-7} \cmidrule(lr){8-11}
    &  &  &  WB-PESQ & NB-PESQ & STOI & SI-SDR & WB-PESQ & NB-PESQ & STOI & SI-SDR \\
    \midrule
    Noisy & - & - & 1.822 & 2.753 & 86.62 & 9.033 & 1.582 & 2.454 &  91.52 & 9.07 \\
    DCCRN-E\cite{hu2020dccrn} & 2020 &  3.7 & - & 3.077 & - & - & - & 3.266 & - & - \\
    Conv-TasNet\cite{convtas} & 2020  & 5.08 & 2.750 & - & - & - & 2.730 & - & - & - \\
    PoCoNet\cite{isik2020poconet} & 2020 & 50  & 2.832 & - & - & - & 2.748 & - & -  & - \\
    DCCRN+\cite{lv2021dccrn+} & 2021  & 3.3 & - & 3.300 & - & - & - & 3.330 & - & - \\
    TRU-Net\cite{choi2021real} & 2021 & 0.38 & 2.740 & 3.350 & 91.29 & 14.87 & 2.860 & 3.360 & 96.32 & 17.55 \\
    CTS-Net\cite{li2021two} & 2021 & 4.99 & 3.020 & 3.470 & 92.70 & 15.58 & 2.940 & 3.420 & 96.66 & 17.99 \\ \midrule
    FullSubNet\cite{hao2021fullsubnet} & 2021 & 5.64 & 3.057 & 3.584 & 92.11 & 16.04 & 2.882 & 3.428 & 96.32 & 17.30 \\
    FullSubNet+\cite{fullsubnetplus}  & 2022 & 8.67 & 3.177 & 3.648 & 93.64 & 16.44 & 3.002 & 3.503 & 96.67 & 18.00 \\
    \model{} & 2022 & 4.21 & \textbf{3.218} & \textbf{3.665} & \textbf{93.93} & \textbf{16.82} & \textbf{3.017} & \textbf{3.513} & \textbf{96.74} & \textbf{18.08} \\
    \bottomrule
    \end{tabular}}
    \end{center}
    \vspace{-0.4cm}
\end{table*}

\section{Experiments}
\subsection{Datasets}
We trained and evaluated \model{} on a subset of the DNS Challenge - Interspeech 2021 dataset with 16 kHz sampling rate, which will be called as DNS Challenge dataset in the following sections.
The clean speech set includes 562.72 hours of clips from 2150 speakers.
The noise dataset includes 181 hours of 65302 clips from over 150 classes.
During training, we used dynamic mixing to simulate speech-noise mixture as noisy speech to make full use of the dataset.
To be specific, 75\% of the clean speeches were mixed with the randomly selected room impulse response (RIR) from openSLR26 and openSLR28 \cite{ko2017study} datasets before the start of each training epoch.
After that, the speech-noise mixtures were dynamically generated by mixing clean speeches and noise at a random SNR between -5 and 20 dB.
The DNS Challenge also provides a publicly available test dataset consisting of two categories of synthetic clips \cite{reddy2020interspeech}, namely without and with reverberations.
Each category has 150 noise clips with a SNR distributed between 0 dB to 20 dB.
We used this test set to evaluate the effectiveness of the model.

\subsection{Training setup and baselines}
We used Hanning window with frame length of 32 ms and frame shift of 16 ms to transform the signals to the STFT domain.
Adam optimizer was used with a learning rate of 1e-3.
For the subband units, we set $n=15$ as in \cite{2020subband}, which means 15 neighbor frequencies are taken on each side of each input frequency bin.
During model training, the input-target sequence pairs were generated as constant-length sequences, with the sequence length set to $T = 192$ frames (approximately 3 s). 

In order to verify the effectiveness of the proposed model, we compared the following models.
All of the models used the same experimental settings as well as learning target (cIRM).

\textbf{FullSubNet:} 
The model consisted of a fullband model and a subband model, each containing two layers of stacked LSTMs and one fully-connected layer.
The fullband model had 512 hidden units per LSTM layer, while the subband model had 384 hidden units per LSTM layer.
In addition, compared with the subband model, the fullband model had an additional ReLU activation function after the fully-connected layer.

\textbf{FullSubNet+:} To further demonstrate the advantages of our model with a small number of parameters and excellent performance, we introduced FullSubNet+ \cite{fullsubnetplus} for comparison.
Following the configuration in \cite{fullsubnetplus}, the number of channels in the MulCA module was 257 and the sizes of the kernels of the parallel 1-D depthwise convolutions were \{3,5,10\} respectively. 
For each fullband extractor, 2 groups of TCN blocks were deployed, each containing 4 TCN blocks with kernel size 3 and dilation rate \{1,2,5,9\}.
The subband model contained 2 LSTM layers with 384 hidden units for each layer.

\textbf{\model{}:} There were 2 groups of TCN blocks deployed in the fullband extractor, each containing 4 TCN blocks with kernel size 3 and dilation rate \{1,2,5,9\}.
The multi-head attention in \camodule{} module had 8 attention heads.
The subband model contained 2 LSTM layers with 384 hidden units per layer.


\begin{table}[!htbp]
    \begin{center}
    \caption{\ninept{Performance of  WB-PESQ [MOS], STOI [\%] and SI-SDR [dB] in the comparative study \ref{secsec:cmp1} using the test set without reverberation.}} 
    \vspace{-0.3cm}
    \label{table:compexp1}
    \scalebox{1.0}{
    \begin{tabular}{cccccc}
    \toprule
    \multirow{2}{*}{Models} &  \multirow{2}{*}{\camodule{}} & \multirow{2}{*}{Concat} & \multirow{2}{*}{\shortstack{WB-\\PESQ}} & \multirow{2}{*}{STOI} & \multirow{2}{*}{SI-SDR} \\
    & &  & & & \\
    \midrule
    \model{}$_a$ & \CheckmarkBold & \XSolidBrush & \textbf{3.017} & \textbf{96.74}  & \textbf{18.08} \\
    \model{}$_c$  & \XSolidBrush & \CheckmarkBold & 2.900 & 96.52  & 17.73 \\
    \model{}$_{ac}$  & \CheckmarkBold & \CheckmarkBold & 2.966 & 96.57 & 17.96 \\
    \bottomrule
    \end{tabular}}
    \end{center}
    \vspace{-0.7cm}
\end{table}

\subsection{Performance Comparison}
Table \ref{table:large total compare} shows the performance of different speech enhancement models on the DNS challenge dataset.
In the table, ``With Reverb" and ``Without Reverb" refer to test sets with and without reverberation respectively.
``\# Para" represents the parameter amount of the model, which is measured in millions.

In the last three rows of Table \ref{table:large total compare}, we compare the performances of FullSubNet, FullSubNet+, and the proposed \model{}.
According to Table \ref{table:large total compare}, \model{} outperforms the baseline FullSubNet in all evaluation metrics with a smaller number of parameters.
In addition, Table \ref{table:large total compare} shows that the FullSubNet+, which is a improved version of FullSubNet, has better performance than FullSubNet. However, the FullSubNet+ also suffers from a large number of parameters and a complex model structure.
In contrast, the \model{} not only outperforms FullSubNet+, but also is achieved with a smaller number of parameters and a simpler architecture.

To further analyse the proposed model, this paper also compares it with other state-of-the-art time domain and frequency domain speech enhancement methods \cite{hu2020dccrn, isik2020poconet, lv2021dccrn+, choi2021real, li2021two} on the DNS Challenge - Interspeech 2021 dataset
in Table \ref{table:large total compare}.
It can be concluded that comparing with the latest methods, the proposed \model{} shows the superior performance on noise reduction tasks without reverberation and even more prominent performance improvement with reverberation. This indicates that the proposed \model{} inherits the excellent capabilities of the FullSubNet for reverberation effects described in \cite{hao2021fullsubnet}, and greatly improves the noise reduction ability by fullband-subband cross-attention module.

\subsection{Investigation of FSCA module}
\label{secsec:cmp1}
In order to investigate fullband-subband cross-attention (FSCA) method, we use \model{} as the backbone and conduct experiments with different fusion methods.
\model{}$_a$ means original \model{} with \camodule{} module for interactive fusion. \model{}$_c$ represents the \model{} with the concatenation between the fullband embedding and subband units as described in FullSubNet \cite{hao2021fullsubnet}, and it does not contain \camodule{} module. \model{}$_{ac}$ denotes the \model{} that uses the above concatenation between fullband embedding and the fusion embeddings outputted by the \camodule{} module. Table \ref{table:compexp1} shows the results of the model on the test set without reverberation in three cases, where ``FSCA" refers to the use of \camodule{} module while ``Concat" refers to the use of the above concatenation approach.

According to Table \ref{table:compexp1}, the performance of \model{}$_a$, \model{}$_{ac}$ are better than \model{}$_c$. 
This indicates that the proposed \camodule{} module can effectively improve the performance of the model in terms of both PESQ and SNR. In addition, the performance of \model{}$_a$ is better than that of \model{}$_c$, which shows that the \camodule{} module can effectively replace the role of the concatenation approach, and reduce the burden of the subsequent subband model by decreasing the input dimension.
Moreover, It can be found that the performance of \model{}$_{ac}$ is not as good as \model{}$_a$.
This is probably because the fusion embeddings has conflicts with the information contained in the original fullband embedding.
\section{Conclusion and Future Work}
This paper proposes a new single-channel speech enhancement framework named \model{}. It adopts a fullband-subband cross-attention (\camodule{}) module to interactively fuse the global and local information.
This paper also deploys a fullband extractor composed of stacked TCN blocks to instead the fullband model in FullSubNet for reducing model size. Experimental results\footnote{Demo page: \href{https://hit-thusz-rookiecj.github.io/FS-CANet}{https://hit-thusz-rookiecj.github.io/FS-CANet}} demonstrate the effectiveness of the fullband-subband cross-attention module.
This paper also compare \model{} with other state-of-the-art methods on the DNS challenge dataset, which shows that the superior performance of proposed \model{}.

This paper only considers incorporating fullband information into the subband information. In the future, we will explore bi-directional fusion of the subband and fullband features.

\textbf{Acknowledgement}: This work is supported by National Natural Science Foundation of China (62076144), Tencent AI Lab Rhino-Bird Focused Research Program (JR202143), CCF-Tencent Open Research Fund (RAGR20210122) and Tsinghua University - Tencent Joint Laboratory.


\bibliographystyle{IEEEtran}

\bibliography{mybib}


\end{document}